\documentclass[a4]{myaa}
\usepackage{txfonts}
\usepackage{natbib}
\usepackage[a4paper,breaklinks]{hyperref}
\bibpunct{(}{)}{;}{a}{}{,}
\usepackage{graphicx}
\usepackage{balance}
\usepackage{latexsym}

\newcommand{\cobold}{\ensuremath{\mathrm{CO}^5\mathrm{BOLD}}}
\newcommand{\linfor}{Linfor3D}
\newcommand{\Teff}{\ensuremath{T_\mathrm{eff}}}

\newcommand{\mD}{\ensuremath{\left\langle\mathrm{3D}\right\rangle}}
\newcommand{\xx}{\ensuremath{\mathrm{1D}_{\mathrm{LHD}}}}
\idline{1}{1}
\begin{document}

\title{The solar photospheric abundance of europium.\subtitle{  
Results from CO5BOLD  3-D hydrodynamical model atmospheres.}}

\author{Alessio Mucciarelli\inst{1}\and
Elisabetta Caffau\inst{2}\and
Bernd Freytag\inst{3}\and
Hans-G\"unter Ludwig\inst{4,2}\and
Piercarlo Bonifacio\inst{4,2,5}}
\authorrunning{Mucciarelli et al.}
\titlerunning{Europium in the solar photosphere}
\offprints{A. Mucciarelli}
\institute{Dipartimento di Astronomia, Universit\`a 
degli Studi di Bologna, Via Ranzani, 1 - 40127
Bologna, Italy
\email{alessio.mucciarelli@studio.unibo.it}\and
GEPI, Observatoire de Paris, CNRS UMR 8111, Universit\'e Paris Diderot;
Place
Jules Janssen 92190
Meudon, France
\email{elisabetta.caffau@obspm.fr, hans.ludwig@obspm.fr, piercarlo.bonifacio@obspm.fr}
\and
Centre de Recherche Astrophysique de Lyon,
           UMR 5574: CNRS, Universit\'e de Lyon,
           \'Ecole Normale Sup\'erieure de Lyon,
           46 all\'ee d'Italie, F-69364 Lyon Cedex 07, France
\email{Bernd.Freytag@ens-lyon.fr}
\and
{CIFIST, Marie Curie Excellence Team}\and
  Istituto Nazionale di Astrofisica,
Osservatorio Astronomico di Trieste,  Via Tiepolo 11,
I-34143 Trieste, Italy
}
\date{Received 25 December 2007; Accepted 6 March 2008}
\abstract  
{Europium is an almost pure r-process element, which
may be useful as a reference in nucleocosmochronology.}
{To determine the photospheric solar abundance using CO5BOLD 
3-D hydrodynamical 
model atmospheres.}
{Disc-centre and integrated-flux observed  solar spectra are used.
The europium abundance is derived from the equivalent width measurements.
As a  reference 1D model atmospheres have 
been used, in addition.}
{The europium photospheric solar abundance is $0.52\pm 0.02$ in agreement
with previous determinations. We also determine the photospheric isotopic
fraction of $^{151}$Eu to be $49\%\pm 2.3\%$ from the intensity spectra and
$50\%\pm2.3\%$ from the flux spectra. This compares well to the
the meteoritic isotopic fraction 47.8\%. 
We explore the 3D corrections also for dwarfs and sub-giants
in the temperature range $\sim$5000\,K to $\sim$6500\,K and 
solar and 1/10--solar metallicities and find them to be negligible for all
the models investigated.}
{Our photospheric Eu abundance
is in good agreement with previous determinations
based on 1D models. This is
in line with our conclusion that 3D effects 
for this element are negligible in the case of the Sun.}
\maketitle

\section{Introduction}   
\label{intro}

Europium (Z=63) is formed through neutron captures
on {\sl seed}-nuclei; in the seminal paper of
\citet{burb} it was assigned both to the {\sl s}-process
(slow neutron capture) and to the 
{\sl r}-process (rapid neutron capture).
According to the current understanding of  
{\sl r}-process nucleosynthesis, europium 
is  an 
almost {\sl pure} {\sl r}-process element. 
About 95-97\% of the Eu
in the solar system is  
contributed by the {\sl r-}process \citep{arl,burris}.
See the introduction in the paper by \citet{MG2000}
on the significance of the Eu/Ba ratio for assessing
the relative contributions of the {\sl r-}process
and {\sl s}-process.

The trend of the [Eu/Fe] abundance ratio as a function of [Fe/H] in 
the Milky Way stars (both halo and disk) seems to mimic the same 
behaviour of the [$\alpha$/Fe] ratio, with an enhanced value in the 
metal-poor stars and a decrease for [Fe/H]$>$-1 dex down to solar 
values. Measurements of this abundance ratio in extra-galactic 
stars highlight a decoupling between [Eu/Fe] and [$\alpha$/Fe] ratios, 
with over-solar [Eu/Fe] at high metallicity and solar or sub-solar 
[$\alpha$/Fe] values, as observed in Sagittarius \citep{boni00}, in 
the Small Magellanic Cloud \citep{hill97} and Large Magellanic Cloud 
\citep{hill95}.

The ratio of the two {\sl r-}process elements Eu/Th
\footnote{The isotope $^{232}$Th is radioactive with a half-life of 14.05 Gyr.}
is potentially an interesting chronometer, provided 
the production ratio of the two nuclei can be reliably
predicted theoretically 
\citep[see][for an extensive review of the {\sl r-}process]{cowan}.

In this paper we reconsider the Eu solar abundance in the light of the 
recent progress of the 3-dimensional model atmosphere computations, 
measuring the solar abundance both with 1D and 3D models
in order to assess the impact of this new generation of models on 
the solar Eu abundance.

The Eu solar abundance reported in the compilation \citet{gs98} 
is  A(Eu)=0.51$\pm$0.08 
\footnote{We adopt the spectroscopic notation A(X)=log(N(X)/N(H))+12.}
and A(Eu)=0.55$\pm$0.02 dex, from the solar photosphere and 
from meteorites, respectively. The analysis 
of \citet{lawler} yields A(Eu)=0.52$\pm$0.01 dex, adopting new atomic data for the 
Eu lines and taking into account the hyperfine structure of the lines employed.

\section{Solar europium chemical analysis}

\subsection{Theoretical tools}

In this paper we derived the solar Eu photospheric abundance by 
using 3D model atmospheres computed with the \cobold\ code 
(COnservative COde for the COmputation of COmpressible COnvection in a 
BOx of L Dimensions with L=2,3) \citep{freytag02,wed} 
and we compared it to the 1D model results.
\cobold\ solves the coupled non-linear equations of compressible 
hydrodynamics including non-local frequency-dependent radiation transport 
for a small volume located at the stellar surface
(for technical issues, see the on-line manual available on
http://www.astro.uu.se/$\sim$bf/co5bold/index.html). 
The atmospheric flow field is sampled in 
equal temporal intervals each of which we call 
a ``snapshot''. In total, 
25 snapshots were selected from a \cobold\ simulation to represent the 
solar photosphere.\\
As reference  we also adopted several 1D 
solar models:

\begin{itemize}
\item  The 1D solar model computed by 
F. Castelli\footnote{http://www.user.oats.inaf.it/castelli/sun/ap00t5777g44377k1asp.dat}
with the ATLAS9 code
adopting the solar abundances of \citet{asplund}, 
\Teff =5777 K, $\log{g}$=4.4377 
and micro-turbulence velocity of 1 km/s.
\item  The semi-empirical solar atmosphere model derived by \citet{hm}. 
\item  A solar 1D atmospheric model  
obtained by the temporal and horizontal average of the 3D structure over surfaces 
of equal Rosseland optical depth. The comparison between the 3D \cobold\ 
model and this kind of {\sl average} 3D model allows to estimate the 
influence of the fluctuations around the mean stratification 
on the line formation process.
\item  A solar 1D atmospheric model computed with a Lagrangian hydrodynamical 
code LHD (see \citealt{caffau07}), including the same opacities and equation-of-state 
adopted by the \cobold\ 3D code. LHD code treats the convection
with the standard mixing length theory (MLT) in the 
formulation given by \citet{mih}. 
The use of this kind of models 
allows to compare directly these 1D models with the 3D \cobold\ models, 
erasing the systematics due to different physical assumptions.
\end{itemize}

The spectral synthesis from ATLAS and HM models  
are performed by using the SYNTHE code \citep{K93,kurucz} in 
its Linux version \citep{sbordone04,sbordone05}.
For \cobold\ and LHD models the
\linfor\footnote{http://www.aip.de/~mst/Linfor3D/linfor\_3D\_manual.pdf}
code is used.

\subsection{Observational material}
The present study is based on two sets of high-resolution, high signal-to-noise ratio 
spectra of solar flux and disc-centre intensity:

\begin{itemize}
\item {\sl Solar Flux --} We adopted the solar flux spectra 
of \citet{neckel84}
and of Kurucz 2005\footnote{See http://kurucz.harvard.edu/sun.html}.
\item {\sl Solar Intensity --} As centre disc solar intensity spectra, 
we used the intensity of \citet{neckel84} and that of 
\citet{deb}\footnote{http://bass2000.obspm.fr/solar\_spect.php}.
\end{itemize}

We selected 5 \ion{Eu}{ii} optical spectral lines from the list of \citet{lawler}.
It is worth to note that not all these solar spectra are useful to measure 
the selected features. 
In the following, we describe briefly the single \ion{Eu}{ii} features used in this
analysis and the corresponding adopted solar spectrum:

\begin{itemize}
\item 412.972 nm -- Strong \ion{Eu}{ii} line, with a weak blending on 
the blue side; this feature exhibits the same shape in all the adopted solar spectra, 
without telluric contamination.
\item 604.951 nm -- Clean feature without particular difficulties, 
it is not blended or contaminated by nearby lines to cause problems in the
placement of the continuum.
\item 664.519 nm -- This is one of the strongest optical transitions and 
is commonly used to infer 
the \ion{Eu}{ii} abundance. In all the four solar observed spectra we consider 
this line results not blended with telluric features.
This feature exhibits in the solar spectrum a 
blending in its red wing, due to the presence of the weak features of 
Cr I and Si I. 
\item 707.709 nm -- This weak line shows an H$_{2}$O telluric line 
contamination on its red side and only in the Delbouille spectrum this blending 
seems to be less severe.
\item 721.756 nm -- This line is measurable only 
in the Delbouille spectrum, instead the other adopted solar spectra 
exhibit a strong blending of this feature with a H$_{2}$O telluric line, 
completely absent in the Delbouille spectrum.
\end{itemize}

\subsection{Chemical analysis}

\subsubsection{Analysis}

The chemical analysis of the selected Eu II features has 
been performed adopting the atomic parameters 
for the \ion{Eu}{ii} lines by \citet{lawler} and summarised 
in Table 1. The \ion{Eu}{ii} spectral lines display 
significant hyperfine structures. We included in the 
line list hyperfine structure and isotopic splitting, 
adopting the meteoritic isotopic ratio 
\footnote{This element has two isotopes, 
with Z=151 and Z=153 with meteoritic 
abundance of 47.8\% and 52.2\% respectively \citep{ag89}} 
and the hyperfine constants A and B measured by \citet{lawler}.
The calculation of the hyperfine structure was done using the 
code LINESTRUC, described by \citet{wal}. 
All the hyperfine components 
for each \ion{Eu}{ii} feature, computed without 
the assumption of a specific isotopic ratio, are
available in the on-line version.
We did not take into account possible 
NLTE effects which are different between intensity and flux spectra 
and could explain the small positive difference between intensity 
and flux abundances: \citet{mash} analysed the NLTE effects for the 
resonance \ion{Eu}{ii} line at 421.9 nm in solar-like stars, finding 
a NLTE correction of $\sim$0.04 dex.

The solar Eu abundance was derived from the curve of growth of each line 
calculated with \linfor, adopting a meteoritic isotopic ratio.  
The equivalent width (EW) of the \ion{Eu}{ii} lines was measured with a Gaussian fit by
using the IRAF task SPLOT, adopting the deblending option. 
The 3D models include only the \ion{Eu}{ii} lines, without the contribution 
of possible blending features. The choice to infer the abundance by using the 
EW measurement comes from the inefficiency of the line profile fitting
with a 3D grid due to the lack of the weak blending components in the 3D
synthetic spectra.
This is due to the inability of the current version of \linfor\ to handle a large
number of lines.
In Table 2  we  provide our results for both 1D and 
3D models, the 3D correction defined by \citet{caffau07} as 
$A(X)_\mathrm{3D}$-$A(X)_\mathrm{\xx}$, and the difference between 
3D and \mD\ models. We reported also the error ($\sigma_\mathrm{EW}$) 
in the Eu abundance due to the uncertainty in the EW measurement (in order to 
estimate this latter issue we performed EW measurements with different 
continuum placements and deblending assumptions for each line), 
typically of $\sim$0.02-0.03 dex (only the 
\ion{Eu}{ii} line at 412.972 nm shows an error in the abundance 
of $\sim$0.05 dex, probably due to the blending on the blue side).

To place solar 3D correction results in a wider context,
we computed 3D corrections of the 664.5 nm \ion{Eu}{ii} line 
for flux spectra in F and G-type atmospheric stellar models.
We explored a parameter grid including \Teff\ between 4980 
and 6460 K, log~g=3.5,4.0,4.5 and [M/H]=0.0,--1.0. 
The Eu abundance is scaled with respect to the metallicity
of the model, according to the solar ratio.
The reference solar Eu abundance is 0.52. The 
results are listed in Table \ref{3dcor}. The majority of the 
3D corrections (3D-\xx) are negligible, and the largest is 
just 0.011\,dex. The 3D correction related to the average 
temperature profile (3D-\mD) is in the range 0.01-0.02\,dex 
for all models and it is larger
than the complete 3D correction.

As additional check to test the consistency of our results, we performed 
a {\sl classical} 1D analysis on these 5 features. 
This step is necessary to compare the results obtained by the LHD models 
and the 1D models usually used in the chemical analysis. 
To compute the abundance we used line profile fitting and employed 
the line list from the Kurucz database, updated including the atomic parameters 
for the \ion{Eu}{ii} lines. 
This was done by using a code \citep{caffau} that 
performs a linear interpolation in a synthetic spectra grid with 
the Eu abundance as a free fitting parameter:
the final best-fit is obtained by 
the numerical $\chi^2$ minimisation,  using MINUIT \citep{james}.
Even the line shift and the continuum placement can be a free
parameter to be adjust to optimise the fit.
Only for the two strong features (namely 412.9 and 664.5 nm) we adopted 
a different version of this code, including as free fitting 
parameters both the Eu abundance and 
the fraction of the Eu isotope $^{151}$Eu with respect 
to the total abundance, $\log(N(^{151}{\rm Eu})/N({\rm Eu}_\mathrm{tot}))$.

\section{Results and discussion}

The main results of this analysis are:
\begin{enumerate} 

\item The 3D analysis, based on different high-resolution high 
signal-to-noise solar spectra and by using the \cobold\ model, provided 
a mean Eu photospheric abundance of A(Eu)=0.506 dex with a standard 
deviation $\sigma$=0.008 for the flux spectra (by using the first three 
spectral features) and A(Eu)=0.527 with $\sigma$=0.024 for the intensity 
spectra at the disk-centre (by using all the five lines). As a final Eu 
solar photospheric abundance we recommend A(Eu)=0.518 dex ($\sigma$=0.024).
This value comes from the average of all the measurements (both flux 
and intensity, since they are very close), and 
it is consistent with the previous 1D determinations.\\  

\item The difference 3D-\mD\  allows to estimate the 
3D corrections due to the horizontal temperature 
fluctuations (a component not taken into account in the classical 
1D models). This correction is negligible for all  of the lines 
considered, with an average value of
--0.009 dex ($\sigma$=0.016) and 0.010 dex ($\sigma$=0.003) for 
flux and intensity respectively. This difference between the two 
solar data-sets has been already observed in a previous 3D analysis 
for sulphur \citep{caffau} and phosphorus \citep{caffau07} and 
can be ascribed to the different atmospheric layers where intensity 
and flux spectra originate (the centre disc intensity spectra 
arise from deeper layers, where the temperature 
fluctuations are more pronounced).\\ 
 
\item The difference 3D-\xx\ allows to compare 
3D and 1D models which employ the same physical assumptions, 
like equation of state and opacities, and provides a 
{\sl 3D correction}.
These 
values are near to zero both   for flux and 
intensity; with an average difference 
of 0.004 dex ($\sigma$=0.013) and 
0.021 dex ($\sigma$=0.004) respectively. 
The 3D-\xx\ corrections appear 
to be systematically higher than  3D-\mD\ corrections  with 
a difference of about 0.010 dex.\\ 

\item Finally, as consistency check, we performed a classical 
1D analysis by using the \citet{hm} and ATLAS models. We derived a 
mean photospheric abundance for europium of A(Eu)=0.515 dex 
($\sigma$=0.022) and 0.523 dex ($\sigma$=0.014) for disc-centre 
intensity and flux, respectively, in good agreement with the 
previous ones by \citet{ag89} of 0.51 dex and by \citet{lawler}
of 0.52 dex. Moreover, also the isotopic ratio 
$(N(^{151}{\rm Eu})/N({\rm Eu}_\mathrm {tot}))$  computed from 
the two strongest Eu lines results in good agreement with the 
meteoritic ratio: 0.49 ($\sigma$=0.023) 
and 0.50 ($\sigma$=0.023) for intensity and flux, respectively.
\end{enumerate}

There is no way to know {\em a priori} if the 3D effects
are important for any given line. A detailed calculation
has to be done in each case.
Solar abundances are widely used as a reference and their
implication goes beyond the pure chemical composition, but
touches field such as helioseismology and solar neutrino production.
The low solar abundances of \citet{asplund} have put some strain
on our understanding of both.
As suggested by \citet{bahcall05}, different measurements of
solar abundances, obtained using different observed
spectra and different solar models, allow a better
estimation of the systematic uncertainties.
In the case of Eu 
we conclude that the 3D effects are 
negligible in the Sun and  solar-like stars.
The scenario is very coherent, when we consider 3D 
of the non-solar models. 3D corrections are negligible for both  
solar and slightly metal-poor models.
Moreover, this is in line with the findings discussed 
by \citet{steffen} that investigated granulation corrections in the 
Sun for several elements. Despite Eu is not included in this 
study, we can compare our results for Eu with their findings 
for Sr. These two elements show very similar line formation properties 
for spectral lines with similar excitation potential and oscillator 
strength. Also for Sr, the corrections are negligible, 
typically between --0.02 and +0.02.
Therefore also for Eu, like for S \citep{caffauS}
and P \citep{caffauP} we conclude that the use of
3D models does not imply a substantial downward
revision of the solar abundances with respect
to what was deduced from the use of 1D models. 

\acknowledgements

We warmly thank the referee, A. J. Sauval, for his useful 
suggestions.  
The authors E.C., H.-G.L., P.B. acknowledge financial
   support from EU contract MEXT-CT-2004-014265 (CIFIST).
   We acknowledge use of the supercomputing centre CINECA,
   which has granted us time to compute part of the hydrodynamical
   models used in this investigation, through the INAF-CINECA
   agreement 2006,2007.

\begin{table*}
\caption{Atomic data for the europium lines considered in this work.}
\begin{tabular}{ccc} 
\hline\hline
Wavelength   & {log~gf}& E.P. \\ 
(nm) &  &  (eV)    \\
\hline 
 412.972& 0.22  & 0.000    \\
 604.951&--0.80  & 1.278    \\
 664.510& 0.12  & 1.379  \\
 707.709&--0.72  & 1.249  \\
 721.756&--0.35  & 1.229 \\
\hline
\noalign{\smallskip}
\end{tabular}
\end{table*}

\begin{table*}
\caption{Solar europium abundances from the adopted observed spectra. Col.(1) 
indicates the used solar spectrum: KF: Kurucz flux, NF: Neckel flux, NI: Neckel 
intensity, DI: Delbouille intensity. Col. (2) is the wavelength of the observed 
lines. Col. (3) is the Equivant Width. Col. (4) and (5) are Eu abundance from the
\cobold\ models and the corresponding uncertainty due to the error of the 
Equivalent Width. Cols. (6)-(9) are the results from the 1D chemical analysis 
by adopting two solar models (HM:\citet{hm}. FC: ATLAS9 solar model by Fiorella 
Castelli). Finally, Cols. (10) and (11) are the 3D corrections. }
\begin{tabular}{ccccccccccccccc} 
\hline \hline 
{Spec} & Wave    & {EW} &
{3D}  & {$\sigma_{\rm EW}$} & {1D}& {$N(^{151}$Eu)/N(Eu$_\mathrm{tot})$} &
{1D} & {$N(^{151}$Eu)/N(Eu$_\mathrm{tot})$} &
{3D-\xx} &
{3D-\mD} \\
 &  & & \cobold\ &  & HM & HM & FC & FC & & \\
  &  nm  &   (pm)   &  (dex)& (dex) & (dex) &  &  (dex) &  &(dex) & (dex)\\
\hline
\hline
KF & 412.972  & 5.620  &  0.509  &  0.045 & 0.535  & 0.54  & 0.526  & 0.51  &	 --0.013     &	 --0.030  \\
NF & 412.972  & 5.652  &  0.513  &  0.045 & 0.527  & 0.49  & 0.513  & 0.48  &	 --0.014     &	 --0.030  \\
NI & 412.972  & 5.026  &  0.537  &  0.050 & 0.533  & 0.51  & 0.537  & 0.49  &	  0.014     &	  0.006  \\
DI & 412.972  & 5.045  &  0.540  &  0.050 & 0.545  & 0.53  & 0.541  & 0.51  &	  0.015     &	  0.006  \\
\hline
KF & 604.951  & 0.063   &  0.504  &  0.030 & 0.528  & ---   & 0.539 &  ---  &	  0.013     &	 0.002    \\
NF & 604.951  & 0.061   &  0.490  &  0.030 & 0.519  & ---   & 0.492 &  ---  &	  0.014     &	 0.003     \\
NI & 604.951  & 0.051   &  0.480  &  0.035 & 0.532  & ---   & 0.487 &  ---  &	  0.023     &	 0.013     \\
DI & 604.951  & 0.057   &  0.529  &  0.035 & 0.537  & ---   & 0.521 &  ---  &	  0.022     &	 0.012     \\
\hline	 						  
KF & 664.519  & 0.436   &  0.511  &  0.020 & 0.540  & 0.52  & 0.520 & 0.51  &	  0.011     &	 0.000     \\
NF & 664.519  & 0.432   &  0.507  &  0.020 & 0.531  & 0.49  & 0.520 & 0.47  &	  0.011     &	 0.000     \\
NI & 664.519  & 0.400   &  0.530  &  0.026 & 0.506  & 0.46  & 0.518 & 0.47  &	  0.020     &	 0.010      \\
DI & 664.519  & 0.384   &  0.514  &  0.021 & 0.504  & 0.50  & 0.504 & 0.49  &	  0.020     &	 0.011      \\
\hline	 						  
DI & 707.709  & 0.082   &  0.526  &  0.032 & 0.480  & ---   & 0.472 & ---  &	  0.025     &	0.012	   \\
\hline	 						  
DI & 721.756  & 0.220   &  0.564  &  0.031 & 0.508  & ---   & 0.513 & ---  &	  0.026     &	0.013	   \\
\hline
\end{tabular}
\end{table*}

\begin{table*}
\caption{The \cobold\ models considered in this work (excluding the solar 
model) for the \ion{Eu}{ii} spectral line at 664.519 nm: the table reports 
the atmospheric parameters (\Teff/$\log{g}$/[M/H]) 
for each model, the EW measurement and the corresponding 3D-\xx\ and 3D-\mD\
corrections.}
\label{3dcor}
\begin{tabular}{lccc} 
\hline\hline
Model parameters & EW  & 3D-\xx  & 3D-\mD   \\ 
(\Teff / log~g / [M/H])& (pm) & (dex) & (dex)   \\
\hline 
  5430/3.5/0.0     & 1.070 & --0.001 &  0.007 \\
  5480/3.5/--1.0 & 0.190 &  0.007 &  0.019 \\
  5930/4.0/0.0     & 0.670 &  0.004 &  0.008 \\
  5850/4.0/--1.0 & 0.110 &  0.009 &  0.023 \\
  4980/4.5/0.0     & 0.400 &  0.000 &  0.009 \\
  5060/4.5/--1.0 & 0.073 &  --0.003 &  0.009 \\
  5870/4.5/0.0     & 0.430 &  0.006 &  0.015 \\
  5929/4.5/--1.0 & 0.066 &  0.011 &  0.021 \\
  6230/4.5/0.0     & 0.400 &  0.009 &  0.017 \\
  6240/4.5/--1.0 & 0.058 &  0.008 &  0.017 \\
  6460/4.5/0.0     & 0.370 &  0.008 &  0.015 \\
  6460/4.5/--1.0 & 0.051 &  0.001 &  0.013 \\
\hline
\noalign{\smallskip}
\end{tabular}
\end{table*} 

\Online

\begin{table*}
\caption{Linelist for the five selected \ion{Eu}{ii} transitions: log~gf, excitation potential and corresponding isotope for 
each hyperfine component are reported.}
\tiny
\label{isoeu}
\begin{tabular}{cccccccccccc} 
\hline\hline
Wavelength   & {log~gf}& E.P. & Isotope   &  Wavelength   & {log~gf}& E.P. & Isotope   &  Wavelength   & {log~gf}& E.P. & Isotope\\ 
(nm) &  &  (eV) &  & (nm) &  &  (eV) &  & (nm) &  &  (eV) &  \\
\hline 
\hline
 412.9627  &  --1.34   & 0.000  &  151 & 604.9532  & --2.36   &  1.278 &  151 &  664.5149    & --1.22 &  1.379 &  151	\\
 412.9623  &  --1.81   & 0.000  &  151 & 604.9531  & --2.83   &  1.278 &  151 &  664.5142    & --2.13 &  1.379 &  151	\\
 412.9649  &  --1.81   & 0.000  &  151 & 604.9529  & --2.83   &  1.278 &  151 &  664.5133    & --3.55 &  1.379 &  151	\\
 412.9645  &  --1.28   & 0.000  &  151 & 604.9528  & --2.30   &  1.278 &  151 &  664.5136    & --1.14 &  1.379 &  151	\\
 412.9640  &  --1.62   & 0.000  &  151 & 604.9526  & --2.64   &  1.278 &  151 &  664.5127    & --1.94 &  1.379 &  151	\\
 412.9676  &  --1.62   & 0.000  &  151 & 604.9524  & --2.64   &  1.278 &  151 &  664.5116    & --3.38 &  1.379 &  151	\\
 412.9671  &  --1.15   & 0.000  &  151 & 604.9522  & --2.17   &  1.278 &  151 &  664.5120    & --1.06 &  1.379 &  151	\\
 412.9665  &  --1.56   & 0.000  &  151 & 604.9519  & --2.58   &  1.278 &  151 &  664.5108    & --1.88 &  1.379 &  151	\\
 412.9710  &  --1.56   & 0.000  &  151 & 604.9516  & --2.58   &  1.278 &  151 &  664.5094    & --3.45 &  1.379 &  151	\\
 412.9704  &  --1.00   & 0.000  &  151 & 604.9514  & --2.02   &  1.278 &  151 &  664.5101    & --0.97 &  1.379 &  151	\\
 412.9698  &  --1.59   & 0.000  &  151 & 604.9510  & --2.62   &  1.278 &  151 &  664.5087    & --1.93 &  1.379 &  151	\\
 412.9753  &  --1.59   & 0.000  &  151 & 604.9506  & --2.62   &  1.278 &  151 &  664.5070    & --3.77 &  1.379 &  151	\\
 412.9746  &  --0.85   & 0.000  &  151 & 604.9502  & --1.87   &  1.278 &  151 &  664.5079    & --0.89 &  1.379 &  151	\\
 412.9739  &  --1.78   & 0.000  &  151 & 604.9496  & --2.80   &  1.278 &  151 &  664.5062    & --2.12 &  1.379 &  151	\\
 412.9803  &  --1.78   & 0.000  &  151 & 604.9492  & --2.80   &  1.278 &  151 &  664.5056    & --0.82 &  1.379 &  151	\\
 412.9796  &  --0.70   & 0.000  &  151 & 604.9486  & --1.72   &  1.278 &  151 &  664.5123    & --1.22 &  1.379 &  153	\\
 412.9681  &  --1.34   & 0.000  &  153 & 604.9514  & --2.36   &  1.278 &  153 &  664.5121    & --2.13 &  1.379 &  153	\\
 412.9678  &  --1.81   & 0.000  &  153 & 604.9515  & --2.83   &  1.278 &  153 &  664.5118    & --3.55 &  1.379 &  153	\\
 412.9690  &  --1.81   & 0.000  &  153 & 604.9514  & --2.83   &  1.278 &  153 &  664.5115    & --1.14 &  1.379 &  153	\\
 412.9688  &  --1.28   & 0.000  &  153 & 604.9515  & --2.30   &  1.278 &  153 &  664.5112    & --1.94 &  1.379 &  153	\\
 412.9684  &  --1.62   & 0.000  &  153 & 604.9515  & --2.64   &  1.278 &  153 &  664.5107    & --3.38 &  1.379 &  153	\\
 412.9701  &  --1.62   & 0.000  &  153 & 604.9515  & --2.64   &  1.278 &  153 &  664.5106    & --1.06 &  1.379 &  153	\\
 412.9698  &  --1.15   & 0.000  &  153 & 604.9515  & --2.17   &  1.278 &  153 &  664.5101    & --1.88 &  1.379 &  153	\\
 412.9694  &  --1.56   & 0.000  &  153 & 604.9515  & --2.58   &  1.278 &  153 &  664.5095    & --3.45 &  1.379 &  153	\\
 412.9715  &  --1.56   & 0.000  &  153 & 604.9514  & --2.58   &  1.278 &  153 &  664.5097    & --0.97 &  1.379 &  153	\\
 412.9712  &  --1.00   & 0.000  &  153 & 604.9513  & --2.02   &  1.278 &  153 &  664.5090    & --1.93 &  1.379 &  153	\\
 412.9709  &  --1.59   & 0.000  &  153 & 604.9511  & --2.62   &  1.278 &  153 &  664.5081    & --3.77 &  1.379 &  153	\\
 412.9733  &  --1.59   & 0.000  &  153 & 604.9510  & --2.62   &  1.278 &  153 &  664.5088    & --0.89 &  1.379 &  153	\\
 412.9731  &  --0.85   & 0.000  &  153 & 604.9508  & --1.87   &  1.278 &  153 &  664.5079    & --2.12 &  1.379 &  153	\\
 412.9730  &  --1.78   & 0.000  &  153 & 604.9503  & --2.80   &  1.278 &  153 &  664.5081    & --0.82 &  1.379 &  153	\\
 412.9756  &  --1.78   & 0.000  &  153 & 604.9500  & --2.80   &  1.278 &  153 & 	     &       &        &     \\
 412.9755  &  --0.70   & 0.000  &  153 & 604.9496  & --1.72   &  1.278 &  153 & 	     &       &        &    \\
  \hline
 \hline
Wavelength   & {log~gf}& E.P. & Isotope   &  Wavelength   & {log~gf}& E.P. & Isotope   &  Wavelength   & {log~gf}& E.P. & Isotope\\ 
(nm) &  &  (eV) &  & (nm) &  &  (eV) &  & (nm) &  &  (eV) &  \\
 \hline
 \hline
 707.7169  & --2.34 & 1.249 &  151  & 721.7564  &  --2.08 &  1.229   & 151 &  &  &  &	 \\
 707.7164  & --2.64 & 1.249 &  151  & 721.7573  &  --2.18 &  1.229   & 151 &  &  &  &	 \\
 707.7154  & --2.17 & 1.249 &  151  & 721.7549  &  --2.63 &  1.229   & 151 &  &  &  &	 \\
 707.7156  & --3.60 & 1.249 &  151  & 721.7559  &  --1.92 &  1.229   & 151 &  &  &  &	 \\
 707.7146  & --2.46 & 1.249 &  151  & 721.7574  &  --1.81 &  1.229   & 151 &  &  &  &	 \\
 707.7132  & --2.00 & 1.249 &  151  & 721.7534  &  --2.59 &  1.229   & 151 &  &  &  &	 \\
 707.7134  & --3.49 & 1.249 &  151  & 721.7550  &  --1.83 &  1.229   & 151 &  &  &  &	 \\
 707.7120  & --2.40 & 1.249 &  151  & 721.7570  &  --1.56 &  1.229   & 151 &  &  &  &	 \\
 707.7104  & --1.86 & 1.249 &  151  & 721.7517  &  --2.74 &  1.229   & 151 &  &  &  &	 \\
 707.7104  & --3.60 & 1.249 &  151  & 721.7537  &  --1.84 &  1.229   & 151 &  &  &  &	 \\
 707.7086  & --2.44 & 1.249 &  151  & 721.7564  &  --1.36 &  1.229   & 151 &  &  &  &	 \\
 707.7066  & --1.73 & 1.249 &  151  & 721.7496  &  --3.10 &  1.229   & 151 &  &  &  &	 \\
 707.7063  & --3.94 & 1.249 &  151  & 721.7523  &  --2.01 &  1.229   & 151 &  &  &  &	 \\
 707.7045  & --2.62 & 1.249 &  151  & 721.7554  &  --1.20 &  1.229   & 151 &  &  &  &	 \\
 707.7024  & --1.61 & 1.249 &  151  & 721.7605  &  --2.08 &  1.229   & 153 &  &  &  &	 \\
 707.7125  & --2.34 & 1.249 &  153  & 721.7602  &  --2.18 &  1.229   & 153 &  &  &  &	 \\
 707.7125  & --2.64 & 1.249 &  153  & 721.7602  &  --2.63 &  1.229   & 153 &  &  &  &	 \\
 707.7116  & --2.17 & 1.249 &  153  & 721.7598  &  --1.92 &  1.229   & 153 &  &  &  &	 \\
 707.7125  & --3.60 & 1.249 &  153  & 721.7594  &  --1.81 &  1.229   & 153 &  &  &  &	 \\
 707.7115  & --2.46 & 1.249 &  153  & 721.7590  &  --2.59 &  1.229   & 153 &  &  &  &	 \\
 707.7106  & --2.00 & 1.249 &  153  & 721.7586  &  --1.83 &  1.229   & 153 &  &  &  &	 \\
 707.7112  & --3.49 & 1.249 &  153  & 721.7578  &  --1.56 &  1.229   & 153 &  &  &  &	 \\
 707.7104  & --2.40 & 1.249 &  153  & 721.7572  &  --2.74 &  1.229   & 153 &  &  &  &	 \\
 707.7093  & --1.86 & 1.249 &  153  & 721.7564  &  --1.84 &  1.229   & 153 &  &  &  &	 \\
 707.7095  & --3.60 & 1.249 &  153  & 721.7553  &  --1.36 &  1.229   & 153 &  &  &  &	 \\
 707.7086  & --2.44 & 1.249 &  153  & 721.7543  &  --3.10 &  1.229   & 153 &  &  &  &	 \\
 707.7078  & --1.73 & 1.249 &  153  & 721.7531  &  --2.01 &  1.229   & 153 &  &  &  &	 \\
 707.7071  & --3.94 & 1.249 &  153  & 721.7514  &  --1.20 &  1.229   & 153 &  &  &  &	 \\
 707.7063  & --2.62 & 1.249 &  153  &		&      &	  &	&  &  &  &    \\
 707.7060  & --1.61 & 1.249 &  153  &		&      &	  &	&  &  &  &    \\
 \hline        			   		  
\noalign{\smallskip}				  
\end{tabular}
\end{table*}


\begin{thebibliography}{}
\bibitem[Anders \& Grevesse(1989)]{ag89}
Anders, E., \& Grevesse, N., 1989, Geochim. Cosmochim. Acta, 53, 197
\bibitem[Arlandini et al.(1999)]{arl}
Arlandini, C., Kappeler, F., Wisshak, K., Gallino, R., Lugaro, M., 
Busso, M., \& Straniero, O., 1999, \apj, 525, 886
\bibitem[Asplund, Grevesse \& Sauval(2005)]{asplund}
Asplund, M., Grevesse, N., \& Sauval, A. J., 2005, ASP, Conf. Ser. 336: Cosmic Abundances 
as Records of Stellar Evolution and Nucleosynthesis, 336, 25
\bibitem[Bahcall et al.(2005)]{bahcall05} Bahcall, J.~N., Basu, 
S., \& Serenelli, A.~M.\ 2005, \apj, 631, 1281
\bibitem[Biemont et al.(1982)]{bkm}
Biemont, E., Karner, C., Meyer, G., Traeger, F., \& zu Putlitz, G., 1982, \aap, 107, 166
\bibitem[Bonifacio(2005)]{bonifacio} Bonifacio, P.\ 2005, Memorie 
della Societ\`a Astronomica Italiana Supplementi, 8, 114 
\bibitem[Bonifacio et al.(2000)]{boni00}
Bonifacio, P., Hill, V., Molaro, P., Pasquini, L., Di Marcantonio, P., \&  Santin, P., 
2000, \aap, 359, 663
\bibitem[Burbidge et al.(1957)]{burb}
Burbidge, E. M., Burbidge, R. R., Fowler, W. A. \& Hoyle, F., 1957, Rev. Mod. Physics, 
29, 547
\bibitem[Burris et al.(2000)]{burris}
Burris, D. L., Pilachowski, C. A., Armandroff, T. E., Sneden, C., Cowan, J. J., 
\& Roe, H., 2000, \apj, 544, 302
\bibitem[Caffau \& Ludwig(2007)]{caffau07}
Caffau, E., \& Ludwig, H.-G., 2007, \aap, 467, L11
\bibitem[Caffau et al.(2005)]{caffau}
Caffau, E., Bonifacio, P., Faraggiana, R., Francois, P., Gratton, R. G., \& 
Barbieri, M., 2005, \aap, 441, 533
\bibitem[Caffau et al.(2007)]{caffauS} Caffau, E., Faraggiana, 
R., Bonifacio, P., Ludwig, H.-G., \& Steffen, M.\ 2007, \aap, 470, 699 
\bibitem[Caffau et al.(2007)]{caffauP} Caffau, E., Steffen, M., 
Sbordone, L., Ludwig, H.-G., \& Bonifacio, P.\ 2007, \aap, 473, L9 
\bibitem[Cayrel et al.(1999)]{cayrel}
Cayrel, R., Spite, M., Spite, F., et al., 1999, \aap, 343, 923
\bibitem[Cowan \& Sneden (2004)]{cowan}
Cowan, J. J., \& Sneden, C., 2004, Carnegie Observatories Astrophysics Series, Vol. 4, 
ed. A. McWilliam and M. Rauch, Cambridge Univ. Press, p.27
\bibitem[Delbouille, Roland \& Neven(1973)]{deb}
Delbouille, L., Roland, G., \& Neven, L., 1973,
Photometric Atlas of the Solar Spectrum from $\lambda\lambda$3000 to 
$\lambda\lambda$10000 Liege: Univ. Liege, Institut d'Astrophysique
\bibitem[Freytag et al. (2002)]{freytag02} Freytag, B., Steffen, 
M., \& Dorch, B.\ 2002, Astronomische Nachrichten, 323, 213
\bibitem[Grevesse \& Sauval(1998)]{gs98}
Grevesse, N., \& Sauval, A. J., 1998, SSRv, 85, 161
\bibitem[Hill et al.(1995)]{hill95}
Hill, V., Andrievsky, S., \& Spite, M., 1995, \aap, 293, 347
\bibitem[Hill(1997)]{hill97}
Hill, V., 1997, \aap, 324, 435
\bibitem[Holweger \& M\"uller(1974)]{hm}
Holweger, H. \& M\"uller, E. A., 1974, Sol. Phys., 39, 19
\bibitem[James(1998)]{james}
James, F., 1998, MINUIT, Reference Manual, Version 94.1, CERN, Geneva, Switzerland.
\bibitem[Lawler et al.(2001)]{lawler}
Lawler, J. E., Wickliffe, M. E., Den Hartog, E. A., \& Sneden, C., 2001, \apj, 563, 1075
\bibitem[{{Kurucz}(1993)}]{K93}
{Kurucz}, R. 1993, SYNTHE Spectrum Synthesis Programs and Line
Data.~Kurucz CD-ROM No.~18.~Cambridge, Mass.: Smithsonian Astrophysical
Observatory, 1993., 18
\bibitem[Kurucz(2005)]{kurucz}
Kurucz, R. L., 2005, Memorie della Societ\`a Astronomica Italiana Supplementi, 8, 14
\bibitem[Mashonkina(2000)]{mash}
Mashonkina, L. I., 2000, ARep, 44, 558
\bibitem[Mashonkina \& Gehren(2000)]{MG2000} Mashonkina, L., 
\& Gehren, T.\ 2000, \aap, 364, 249 
\bibitem[Mihalas(1978)]{mih}
Mihalas, D., 1978, Stellar atmospheres, 2nd edition (San Francesico, 
W. H.Freeman and Co., 1978)
\bibitem[Neckel \& Labs(1984)]{neckel84}
Neckel, H., \& Labs, D., 1984, Sol. Phys., 90, 205
\bibitem[{{Sbordone} (2005)}]{sbordone05}
{Sbordone}, L. 2005, Memorie della Societ\`a Astronomica Italiana Supplementi, 8, 61
\bibitem[{{Sbordone} {et~al.} (2004){Sbordone}, {Bonifacio}, {Castelli}, \&
  {Kurucz}}]{sbordone04}
{Sbordone}, L., {Bonifacio}, P., {Castelli}, F., \& {Kurucz}, R.~L. 2004,
  Memorie della Societ\`a Astronomica Italiana Supplementi, 5, 93
\bibitem[Steffen \& Holweger(2002)]{steffen}
Steffen, M, \& Holweger, H., 2002, \aap, 387, 258
\bibitem[Wahlgren(2005)]{wal}
Wahlgren, G. M., 2005, Memorie della Societ\`a Astronomica Italiana Supplementi, 8, 108
\bibitem[Wahlgren et al.(1995)]{wal95}
Wahlgren, G. M., Leckrone, D. S., Johansson, S. G., Rosberg, M., \& Brage, T., 1995, \apj, 444, 438
\bibitem[Wedemeyer et al. (2004)]{wed} Wedemeyer, S., 
Freytag, B., Steffen, M., Ludwig, H.-G., \& Holweger, H.\ 2004, \aap, 414, 
1121 
\end{thebibliography}
\end{document}